\documentstyle[aps,epsf,draft,preprint]{revtex}

\begin{document}
\author{Luke Rallan$^{*}$, and Vlatko Vedral$^{\dagger}$}
\title{Energy Requirements for Quantum Data Compression and 1-1 Coding}
\address{$^{*}$ Centre for Quantum
Computation, Clarendon Laboratory,
    University of Oxford,
    Parks Road,
    Oxford OX1 3PU, United Kingdom\\
$^{\dagger}$ Optics Section, The Blackett Laboratory, Imperial College, Prince Consort
Road, London SW7 2BZ, United Kingdom}
\date{\today}
\maketitle

\begin{abstract}
By looking at quantum data compression in the second
quantisation, we present a new model for the efficient generation
and use of variable length codes. In this picture lossless data
compression can be seen as the {\em minimum energy} required to
faithfully represent or transmit classical information contained
within a quantum state.

In order to represent information we create quanta in some
predefined modes (i.e. frequencies) prepared in one of two
possible internal states (the information carrying degrees of
freedom). Data compression is now seen as the selective
annihilation of these quanta, the energy of whom is effectively
dissipated into the environment. As any increase in the energy of
the environment is intricately linked to any information loss and
is subject to Landauer's erasure principle, we use this principle
to distinguish lossless and lossy schemes and to suggest bounds
on the efficiency of our lossless compression protocol.

In line with the work of Bostr\"{o}m and Felbinger
\cite{bostroem}, we also show that when using variable length
codes the classical notions of prefix or uniquely decipherable
codes are unnecessarily restrictive given the structure of
quantum mechanics and that a 1-1 mapping is sufficient. In the
absence of this restraint we translate existing classical results
on 1-1 coding to the quantum domain to derive a new upper bound
on the compression of quantum information. Finally we present a
simple quantum circuit to implement our scheme.
\end{abstract}

\section{Introduction}
Data compression is already a fundamental and well developed
branch of classical information theory.  It has wide reaching
implications on every aspect of information storage and
transmission and its quantum analogue is of considerable interest
in a wide range of applications \cite{review}.  In quantum
information theory the idea of quantum data compression, in its
strictest sense, has still much to gain from the classical theory
with only some of the more fundamental classical notions being
translated \cite{bostroem,schu2,jozsa1,chuang,brau}.

The basis for compression of classical data is Shannon's
noiseless coding theorem \cite{shannon}, which states that the
limit to classical data compression is given by the Shannon
entropy. Schumacher \cite{jozsa1} presented the quantum analog to
this and proved that the minimum resources necessary to
faithfully describe a quantum message in the asymptotic limit is
the von Neumann entropy of the message $\rho$, given by $S(\rho)
= - tr \rho \log \rho$. Schumacher also demonstrated that by
encoding only the typical subspaces this bound could be achieved
using a fixed length block coding scheme, and how in the
asymptotic limit the compressed message can be recovered with
average fidelity arbitrarily close to unity.

The fact that Schumacher's scheme is only faithful in the
asymptotic limit has led many to ask whether there is a scheme
where we can losslessly compress in the finite case i.e. where we
want to be able to compress without loss of information in the
case where we have a finite (i.e. more practical) number of
qubits. Of course there are many reasons why such a scenario would
be desirable, e.g. in a quantum key distribution (QKD) scheme
where often very high fidelity of the finite received signal is
crucial to the integrity of the scheme \cite{nielsen}. It is in
such cases that the asymptotically faithful Schumacher compression
would not be ideal.  This has, directly or indirectly, inspired a
number of other quantum schemes \cite{bostroem,chuang,brau} based
on classical ideas of lossless coding, such as Huffman, arithmetic
and Shannon Fano coding \cite{cover}.

The primary consideration in lossless compression schemes is the
efficient generation and manipulation of variable length codes
(i.e. codes of variable rather than fixed block length). This is
because (as proven in particular by Bostr\"{o}m and Felbinger
\cite{bostroem}) it is not possible to achieve truly lossless
compression using block codes. The application of variable length
codes to quantum data compression is however not quite so
straightforward.  The main issue seems to be that we are forbidden
by quantum mechanics to measure the length of each signal without
disturbing and irreversibly changing the state and resulting
message.  In our scheme however we show both compression and
decompression to be unitary operations and there never be any need
for a length measurement of the variable length states.

The main point of this paper is two fold, by looking at quantum
data compression in the second quantisation, we present an
entirely new model of how we can generate and efficiently use
variable length codes. The significance of this model is that we
believe it is a more natural application of variable length
coding in quantum information theory. More importantly still is
that fact that any data compression (lossless or lossy) can be
seen as the {\em minimum energy} required to faithfully represent
or transmit classical information contained within a quantum
state. This allows us to use energy and entropy arguments to give
a deeper insight into the physical nature of quantum data
compression and to suggest bounds on the efficiency of our
lossless compression protocol in a novel and interesting way.

The rest of this paper is broken down as follows;  Section II of
our paper is dedicated to reviewing the second quantisation and
introducing our notation for the description of quantum states.
In this description the average length of the codeword is related
to the number of ``modes" that are occupied. Using this fact, we
look at the average energy of the message instead of its average
length and therefore represent the compression limit from an
energy rather than length perspective. In Section III this offers
the possibility to interpret data compression in terms of the
Landauer erasure principle.  In Section IV we introduce a
compression algorithm that uses the second quantisation to
generate variable length codes and show how the need for prefix
or uniquely decipherable codes is unnecessarily restrictive given
the structure of quantum mechanics. The absence of this restraint
leads us to the concept of one-to-one ('1-1') codes. Classical
results are then used to present an analogous quantum 1-1 entropy
bound which, when taking into account the classical side
information, asymptotically tends towards the existing Von
Neumann bound.  Finally, in Section V, we give an experimental
setup for a small example that could be used to demonstrate the
legitimacy of this new compression algorithm.

\section{Energy and Coding}
In this section we introduce our second quantisation notation and
then show, initially using Schumacher's scheme as an example, how
data compression can be seen as the {\em minimum energy} required
to faithfully represent or transmit classical information
contained within a quantum state.

The general scenario for data compression is that a memoryless
source, say Alice, wants to send a message $M$ to a receiver Bob,
in the most 'efficient' way possible. The efficiency of this
communication in space or time may be described through the
optimisation of any one of a number of parameters e.g. minimising
the number of bits or the total energy required to represent the
message (the two are not necessarily equal). The scenario we use
in this paper is similar to that employed by Schumacher
\cite{schu1}. In our protocol the source Alice, wishes to
communicate a number, $n$, of quantum systems (which we call the
{\em letters}) prepared from a set of $N$ distinct (but not
necessarily mutually orthogonal) states $|\psi_1\rangle,
|\psi_2\rangle \ldots |\psi_N\rangle$ to Bob in the most
efficient manner. It is also worth clarifying that like in
Schumacher's scheme, we consider the compression of a single
source message, this is in contrast to other many message schemes
with an extended memory/source set \cite{bostroem}. In addition
in our scheme our objective is to minimise the energy of any given
sequence of states generated by the source having only the
probabilities of the source and indeed knowledge of the letter
states themselves. Here each letter corresponds to a distinct part
of the message, with $n$ letters representing the whole message.
Alice then compresses this message and sends the statistical
properties to Bob. Bob then adds the redundancy back into the
message (using the classical side channel, as we see later) and
then performs a set of transformations to determine the correct
sequence of states comprising Alice's letters, which he can then
begin to use to reconstruct the entire message. This last part is
known as decompression of the message. In this paper, we assume
that the communication is noiseless, i.e. the states suffer no
error on the way to the Bob, the letters are statistically
independent from one another (i.i.d.), and for comparison with
other classical and quantum compression schemes and without loss
of generality that Alice communicates qubits.

There are a number of physical systems that may be used to
realise qubits e.g. two different polarisations of a photon, two
different possibilities for the alignment of nuclear spins in a
uniform magnetic field (``up and down"), or the two energy levels
of an electron orbiting say a hydrogen atom. In this paper,
although our terminology for encoding and manipulating
information refers to polarisations of a photon, the underlying
theory and results can be conveniently applied to other qubit
realisations (matter or field alike).

In addition to polarisation ($P$) let us use another degree of
freedom, say frequency ($\omega$). A third possible degree of
freedom is spatial location or coordinate, which will be used in
section VI. Now in the second quantisation picture with the set of
frequencies $\omega_i$, and the polarisations, $H$ or $V$ (i.e.
whether the photon horizontally or vertically polarised) we can
represent a system consisting of a variable number of photons by
the following basis states:

\begin{eqnarray}
|\psi_{n_1,...n_N;m_1...m_N}\rangle = |n{{_1}_H},
m{{_1}_V}\rangle_{w_1}\ldots |n{{_N}_H}, m{{_N}_V}\rangle_{w_N}
\nonumber
\end{eqnarray}

Here we have $N$ different modes of the system i.e. $N$ different
frequencies, $w$. Each mode is made up of 2 different harmonic
oscillators, one for each polarization.  It is worth noting that
in our protocol these frequencies do not act as additional {\em
information carrying} degrees of freedom, they are merely
placeholders to distinguish the qubits and are assumed to fixed
apriori by the source or sender and receiver (a good example of
this is if we consider the normal modes in a cavity). That these
prior correlations exist between the sender and receiver is a
common and necessary part of any spatial or temporal information
transfer.

Anyhow writing the system in this basis tells us that we have:
$n_1$ photons with frequency $w_1$ in a horizontal polarisation,
$m_1$ photons with frequency $w_1$ in a vertical polarisation and
so on until $n_N$ photons with frequency $w_N$ in a horizontal
polarisation and $m_N$ photons with frequency $w_N$ in a vertical
polarisation. Note that states with different number of photons
are orthogonal. The most general state of this system is a
superposition of all the basis states above:

\begin{eqnarray}
|\psi\rangle = \hspace{-0.3cm} \sum_{{n_1}{m_1}\ldots {n_N}{m_N}}
\hspace{-0.5cm} a_{{n_1}{m_1}\ldots {n_N}{m_N}} |n{{_1}_H},
m{{_1}_V}\rangle_{w_1}\ldots |n{{_N}_H}, m{{_N}_V}\rangle_{w_N}
\nonumber
\end{eqnarray}

In practice there are infinitely many modes, but we only consider
the ones which are occupied.  The unoccupied modes will be said
to be in a vacuum state. If our state consisted of say only $1$
vertically polarised photon in only the first mode, this could be
represented as:

\begin{eqnarray}
|\psi\rangle = |0_H,1_V \rangle _{w_1} |0_H,0_V \rangle _{w_2}
|0_H,0_V \rangle _{w_3} ........... |0_H,0_V \rangle _{w_\infty}
\nonumber
\end{eqnarray}

All of the states can be generated from the overall vacuum state
or zero state (which is the one containing vacuum in every mode).
They can be generated (or destroyed) by applying creation (and
annihilation) operators, $a$ (and $a^{\dagger}$) respectively,
where there are separate operators for both horizontally and
vertically polarised photons.  In general we stick to only one
excitation per mode. In addition given that these modes are fixed
apriori by the source or sender and receiver, there is exactly
one photon in all the modes between the lowest frequency mode and
the first vacuum mode.  We keep to one excitation per mode to
ensure that this scheme applies equally well to both bosonic and
fermionic systems and keeping the scheme as universal as possible.

In this framework it is easier to see that compressing from $n$
qubits to $m$ qubits leads to a reduction in average energy. This
is because in order to create or annihilate one photon in the
mode $w_i$, requires investing or releasing an amount of energy
equal to $\hbar \omega_i$. To see this, let us first write the
Hamiltonian for this system:

\begin{eqnarray}
H = \sum_n \hbar w_n ({{a^{\dagger H}}_n} {{a_n}^H +
{a^\dagger}_n}^V {a_n}^V) \nonumber
\end{eqnarray}
which is the standard harmonic oscillator Hamiltonian for every
mode and polarization, summed up over all of them independently.
In this Hamiltonian the modes are independent and
non-interacting. The interaction between the modes will be added
later and used for data compression and decompression.

Suppose now that a quantum source randomly prepares different
qubit states $|\psi_i\rangle$ with corresponding probabilities
$p_i$ (keeping this analysis general we can apply this to systems
of higher dimension than qubits). A random sequence of $n$ such
states is produced. The question is, by how much can this be
compressed, i.e. how many qubits do we really need to encode the
original sequence (in the limit of large $n$)? First of all the
total density matrix is
\begin{eqnarray}
\rho = \sum_i p_i |\psi_i\rangle\langle \psi_i| \nonumber
\end{eqnarray}

After the compression the total message would consist of some
smaller number $m$ of qubits, in the state $\rho_{m}$ and it is
the ratio ${m}/{n}$ that our compression aims to minimise. In the
case of Schumacher's compression this is achieved by projecting
$\rho_{n}$ onto the typical subspace and then if the projection is
successful the resulting strings are then encoded using a block
coding scheme analogous to that employed by Shannon
\cite{shannon}.  In the asymptotic case the probability that we
are not successful in this projection goes exponentially close to
zero as $n$ increases.  Therefore the efficiency of quantum
encoding is the same as the efficiency of the classical block
coding scheme used, after the successful projection onto the
typical subspace. The best way of deriving this is to look at the
density matrix in the diagonal form

Now, this matrix can be diagonalised
\begin{eqnarray}
\rho = \sum_i r_i |r_i\rangle\langle r_i| \nonumber
\end{eqnarray}
where $r_i$ and $|r_i\rangle$ are the eigenvectors and
eigenvalues. This decomposition is, of course, indistinguishable
from the original one (or any other decomposition for that
matter). Thus we can think about compression in this new basis,
which is easier as it behaves completely classically (i.e. they
are fully distinguishable since $\langle
r_i|r_i\rangle=\delta_{ij}$). We can therefore invoke results
from Shannon's work on classical typical sequences \cite{shannon}
to conclude that the limit to compression is $n(-\sum_i r_i \ln
r_i)$, i.e. $n$ qubits can be encoded into $nS(\rho)$ qubits. No
matter how the states are generated, as long as the total state
is described by the same density matrix $\rho$ its compression
limit is its von Neumann entropy.

We now want to show briefly that this is the same as the ratio of
the initial and final energy in the second quantisation and data
compression is therefore the same as energy reduction in our
communication framework.  The average initial (before
compression) and final (after compression) energies in our
message are defined by the usual trace rule:

\begin{eqnarray}
\langle H \rangle_{initial} = tr(H\rho_{initial}) \nonumber \\
\langle H \rangle_{final} = tr(H\rho_{final}) \nonumber
\end{eqnarray}

Asymptotically we claim that:

\begin{eqnarray}
{\langle H \rangle_{final} / \langle H \rangle_{initial}} =
S(\rho)\nonumber
\end{eqnarray}

We now substantiate this claim and a slight change in notation
will make this easier to see.  We define a message of length $n$
to be described by the density matrix $\rho_n =
\rho_{i}^{\otimes{n}} = (\sum_{i=1}^{n}p\left( \psi_{i}\right)
\left| \psi_{i}\right\rangle \left\langle
\psi_{i}\right|)^{\otimes n}$, which again when written in the
diagonal basis gives:

\begin{eqnarray}
\rho_{n} = (\sum_i r_i |r_i\rangle\langle r_i|)^{\otimes{n}}
\nonumber
\end{eqnarray}

this is equal to
\begin{eqnarray}
\rho_{n} = \sum_{i_1,i_2 \ldots i_n} {r_{i_1}, r_{i_2} \ldots
r_{i_n}} |r_{i_1}, r_{i_2} \ldots r_{i_n} \rangle \langle
r_{i_1}, r_{i_2} \ldots r_{i_n}| \nonumber
\end{eqnarray}

For large $n$, the number of times the ket $|r_j\rangle$ appears
is, as a consequence of the law of large numbers, $n r_j$.

Without loss of generality this proof looks at a source, Alice,
generating photons in only one of two states at a time, $|\psi_1
\rangle$ or $|\psi_2 \rangle$ where the state $|\psi_1 \rangle =
a|H\rangle + b|V\rangle$ and $|\psi_2 \rangle = a|H\rangle -
b|V\rangle$. Each generated photon corresponds to a letter of the
message, where the whole length of the message is measured in
qubits. The overlap between the two states is $a^2-b^2$ and if
$b=\sin (\theta/2)$ and $a=\cos (\theta/2)$ this overlap becomes
$\cos\theta$. Therefore our two eigenstates are:
\begin{eqnarray}
|r_1\rangle = \frac{1}{\surd{2}}(|H\rangle + |V\rangle)
\nonumber\\
|r_2\rangle = \frac{1}{\surd{2}}(|H\rangle - |V\rangle)\nonumber
\end{eqnarray}

which when looked at in the second quantisation framework can be
rewritten as:

\begin{eqnarray}
|r_1\rangle & = & \frac{1}{\surd{2}}(|1_H, 0_V\rangle + |0_H, 1_
V\rangle) \nonumber\\
|r_2\rangle & = & \frac{1}{\surd{2}}(|1_H, 0_V\rangle - |0_H, 1_
V\rangle) \nonumber
\end{eqnarray}

A compression-decompression scheme of rate $R$ consists of two
quantum operations ${C}'$ and ${D}'$ analogous to the maps
defined for the classical case \cite{cover} where the compression
operation ${C}'$ is taking states from $H^{\otimes n}$ to
$H^{\otimes nR}$ and the
decompression ${D}'$ returns them back. One can define a sequence $%
r_{i_{1}}r_{i_{2}}r_{i_{3}}\cdots r_{i_{n}}$ as $\epsilon
$-typical by a relation similar to the classical

\begin{eqnarray}
\left| \frac{1}{n}\log \left( \frac{1}{\left( r_{i_{1}}\right) \left( r_{i_{2}}\right)
\cdots \left( r_{i_{n}}\right) }\right) -S\left( \rho \right) \right| \leq \epsilon .
 \nonumber
\end{eqnarray}

This is a rigorous statement of the law of large numbers
\cite{cover}.  A state $\left| r_{i_{1}}\right\rangle \left|
r_{i_{2}}\right\rangle \cdots \left| r_{i_{n}}\right\rangle $ is
said to be $\epsilon $-typical if the
sequence $r_{i_{1}}r_{i_{2}}r_{i_{3}}\cdots r_{i_{n}}$ is $\epsilon $%
-typical. The $\epsilon $-typical subspace will be noted $T\left(
n,\epsilon\right) $ and the projector onto this subspace will be:

\begin{eqnarray}
P(n,\epsilon)={r_{i_{1}} \cdots r_{i_{n}} \in T(n,\epsilon)} {\sum }\left|
r_{i_{1}}\right\rangle \left\langle r_{i_{1}}\right| \otimes \cdots \otimes \left|
r_{i_{n}}\right\rangle \left\langle r_{i_{n}}\right|\nonumber
\end{eqnarray}

By using a block coding scheme to encode only the typical subspace, Schumacher manages
to compress the message from $n$ to $m$ qubits, so that the state is:

\begin{eqnarray}
\rho_{m} & = &  \hspace{-0.1cm} \sum_{i_1,\ldots i_m} t_{i_1}
\ldots t_{i_m} |t_{i_1} \ldots t_{i_m} \rangle \langle t_{i_1}
\ldots t_{i_m}| \otimes (|0_{m+1} \ldots 0_n \rangle \langle
0_{m+1} \ldots 0_n|) \nonumber
\end{eqnarray}

Writing this in our representation, this means that only the
first $m$ modes are populated, the other $n-m$ modes being empty.
Schumacher proved that in the asymptotic limit the average number
of qubits required tends to the Shannon entropy. Therefore from
our energy perspective we can say that, as originally each qubit
corresponds directly with a mode containing a photon, on average
we are saving $n-m$ photons. Assuming all photons have the same
energy of $\hbar \omega$ (i.e. all the modes have the same
frequency and energy is the same for both $H$ and $V$
polarisations) then the following is true:

\begin{eqnarray}
\langle H \rangle_{initial} = \hbar \omega & n \nonumber \\
\langle H \rangle_{final} = \hbar \omega & m \nonumber
\end{eqnarray}
With regards to the assumption that all photons are of the same
energy, for this proof we can for example practically assume that
the frequencies are very close to each other (emitted from a large
cavity). Not only that, but we can also assume that we always emit
a photon from the same mode - but then the spatial component
(equivalently the temporal component) of all these would be
different which is how we could discriminate them. Anyhow
following on from this, therefore:
\begin{eqnarray}
\frac{\langle H \rangle_{final}} { \langle H \rangle_{initial}} =
\frac{m}{n}\nonumber
\end{eqnarray}

From Schumacher, the ratio $m/n$ we know to be equal to
$S(\rho_{n})$.  The entropy therefore tells us not only how much
information can be compressed but also how much energy is used
for computation. In the Schumacher block coding scheme
\cite{schu1} the average message length divided by the final
message length gives us the entropy. Given that for us the
average message length is the number of kets that are non zero, it
makes more sense to look at average energy rather than average
length. In this way we reinterpret the entropy from an energy
rather than length perspective.

Note that here we used Schumacher's compression to illustrate how
compression works, but any other compression scheme would also
amount to energy reduction.  In fact in section IV we will present
a scheme completely different to Schumacher's (i.e. faithful for
any finite length of message) which can also be regarded as
reducing the energy of the original message whilst preserving its
information content.  A conceptual advantage of our formulation
is that we have a more physical interpretation of data
compression through Landauer's erasure, which we look at next.

\section{Landauer Erasure Perspective}
We can also show how Landauer's principle \cite{landauer} can be
used to derive the limit to quantum data compression.  This
principle states that in order to delete a message containing
entropy S, it is necessary to use kT S energy.  In particular
this means that in order to delete 1 bit of information from some
message you need to generate at least one bit of entropy (heat)
in the environment of the message. In order to see this suppose
that the qubit to be deleted is in the maximally mixed state
$I_m/2$. We know there is no unitary (reversible) transformation
that maps this into a pure state. The best that we can do is swap
a clean pure qubit from the environment with the maximally mixed
one to be deleted \cite{vedral2}.

\begin{eqnarray}
U_s (I_m\otimes|0_e\rangle\langle0_e|) U_s^{\dagger}= |0_m\rangle \langle
0_m|\otimes\frac{I_e}{2}\nonumber
\end{eqnarray}
where the swap transformation is defined as:
\begin{eqnarray}
U_s(|\psi_m\rangle|\phi_e\rangle)= |\psi_e\rangle|\phi_m\rangle\nonumber
\end{eqnarray}

In this way the entropy of $1$ (qu)bit that existed in the system
is now transferred to the environment as since the whole
transformation is at best unitary (reversible) we can see that
the environment cannot increase in entropy by less than $1$
(qu)bit. This is the key observation of Landauer's principle. If
the qubit that we are erasing is originally entangled to another
qubit of the message, then after the swap operation with the
environment, the corresponding environmental qubit becomes
entangled with the message (This is the simplest instance of the
so called entanglement swapping scheme and will be exploited in
our later implementation of our own data compression algorithm
scheme). Note that there are several ways of formulating this
principle. It could be phrased in terms of entropy or in terms of
free energy or even in terms of the heat that is generated. These
are all equivalent. All this thermodynamical reasoning applies in
the so-called thermodynamical limit, i.e when we manipulate a
large number of systems (in our case this means a sufficiently
long message).

The fact that we release energy but do not lose any information is
fundamental to lossless quantum data compression. By controlling
the release of 'redundant energy', that carries no information we
can keep the quantum coherences intact. If no information is lost
during compression the two free energies (or entropies) before
and after the compression should be equal (c.f. \cite{feynman}).
If we generated less heat after compression, this means that
there was less information to delete - implying that our
compression was not faithful (and vice-versa). Therefore if $n$
qubits in a state $\rho$ are compressed to $m$ qubits in a state
$\sigma$ the most efficient compression according to Landauer's
erasure would result in $nS(\rho)= mS(\sigma)$.  In order to
minimise $m/n$ (i.e. achieve the most efficient compression) the
entropy of the encoded bits should be maximal, $S(\sigma)=1$
implying that $\sigma$ has to be in a maximally mixed state.
Since we know Landauer principle has to be obeyed, if the
compressed message generates a lower amount of heat when deleted
this implies that in order to decompress this message we need
another piece of information about it to make up the difference
between Landauer's heat and this heat.  We will see an example of
this in the next section when we introduce our own coding scheme.

\section{1-1 Quantum Coding}

We have seen that Schumacher's encoding is lossy for a finite
length of message and that it only becomes faithful in the
asymptotic limit. Now we present an encoding which is faithful
for any number qubits. Our encoding compresses the quantum
information in the qubits beyond the von Neumann limit, but at
the expense of having to send an additional piece of (classical)
information about the state from source. We will see how when
adding both pieces of information together the efficiency of our
compression scheme, like Schumacher's, tends asymptotically to
the Von Neumann limit. To see how this can be achieved, suppose
again that a quantum source randomly prepares different qubit
states $|\psi_i\rangle$ with the corresponding probabilities
$p_i$ (keeping this analysis general we can apply this to systems
of higher dimension than qubits). A random sequence of $n$ such
states is produced. The question is then how many qubits do we
really need to encode the original sequence i.e. by how much can
the source be compressed?

As before the single-shot density matrix for the source is:
\begin{eqnarray}
\rho_{i} = \sum_i p_i |\psi_i\rangle\langle \psi_i| \nonumber
\end{eqnarray}

If we assume we know the probability distribution $p_i$ (for all $i$) of the source
then this matrix can be diagonalised to give
\begin{eqnarray}
\rho_i = \sum_i r_i |r_i\rangle\langle r_i| \nonumber
\end{eqnarray}

where $r_i$ and $|r_i\rangle$ are the eigenvectors and
eigenvalues. The advantage of diagonalisation is that compression
in this new basis is easier as the state behaves completely
classically (since $\langle r_i|r_i\rangle=\delta_{ij}$). We can
then invoke results on classical compression methods to compress
the resulting state.  The important difference here to other
compression schemes with an extended memory/source set
\cite{bostroem} is that by considering compression in the second
quantisation we have a novel framework for implementation of
variable length codes. We will first introduce our scheme through
an example and then discuss its generalization and efficiency.

\noindent {\bf Example}. \\ Suppose that Alice wishes to generate
and send a classical string of 3 bits (of course our scheme also
generalises to quantum information). Alice encodes each of her
bits into either state $|\Psi_0\rangle = \cos (\theta/2)
|0\rangle + \sin (\theta/2) |1\rangle $ or $|\Psi_1\rangle = \sin
(\theta/2) |0\rangle + \cos (\theta/2) |1\rangle$ with
$p_0=p_1=1/2$ as in Fig. 2 (we assume equal probabilities for
simplicity, however, there is no loss of generality).  The
question is what level of compression can Alice expect to achieve?

The interesting thing here is that classically it is not possible
to compress a source that generates $0$ and $1$ with equal
probability. Quantum mechanically, however, compression can be
achieved not only by the nature of the probability distribution
but also due to the non-orthogonality of the states encoding
symbols of the message. In this example the overlap between the
two states is $\langle \psi_0|\psi_1 \rangle = \sin \theta$ and
they are orthogonal when $\theta = \pi$. Essentially the smaller
the overlap, the more the total message can be compressed. This
can also be expressed in terms of information, as the smaller the
overlap, the less distinguishable the states and hence the less
information they carry.

As Alice's message is only $3$ qubits long. Then there are $8$
different possibilities, $|\Psi_0\Psi_0\Psi_0\rangle,...
|\Psi_1\Psi_1\Psi_1\rangle$, which are all equally likely with
$1/8$ probability.

It is always possible to go to a basis where the density matrix is diagonal (here we
are using the first quantisation notation for clarity):

\begin{eqnarray}
\rho = \frac{(1+\sin \theta )}{2} |+\rangle\langle +| + \frac{
(1-\sin \theta )}{2}|-\rangle\langle -| \nonumber
\end{eqnarray}
where $|\pm\rangle = |0\rangle \pm |1\rangle$. Consequently for 3
qubits:
\begin{eqnarray}
\rho_3 = \rho ^{\otimes{3}} & = & [{\frac{(1+\sin \theta )}{2}
}]^3|+++\rangle\langle +++|  +  [{\frac{(1+\sin \theta
)}{2}}]^2[{\frac{(1-\sin \theta )}{2}}]^1
|++-\rangle\langle -++| \ldots \nonumber\\
& + & [{\frac{(1-\sin \theta )}{2}}]^3
|---\rangle\langle
---| \nonumber
\end{eqnarray}
By rewriting the combined state in the second quantisation and
making use of the fact that the only requirements on the
resulting codeword is that it is unique (i.e. 1-1) we can encode
our message as follows:
\begin{eqnarray}
a^3b^0|+++\rangle & \rightarrow & a^3b^0|0_H,1_V \rangle _{w_1}                                                 \nonumber\\
a^2b^1|++-\rangle & \rightarrow & a^2b^1|1_H,0_V \rangle _{w_1}                                                 \nonumber\\
a^2b^1|+-+\rangle & \rightarrow & a^2b^1|0_H,1_V \rangle _{w_1} |0_H,1_V \rangle _{w_2}                         \nonumber\\
a^2b^1|-++\rangle & \rightarrow & a^2b^1|0_H,1_V \rangle _{w_1} |1_H,0_V \rangle _{w_2}                         \nonumber\\
a^1b^2|+--\rangle & \rightarrow & a^1b^2|1_H,0_V \rangle _{w_1} |0_H,1_V \rangle _{w_2}                         \nonumber\\
a^1b^2|-+-\rangle & \rightarrow & a^1b^2|1_H,0_V \rangle _{w_1} |1_H,0_V \rangle _{w_2}                         \nonumber\\
a^1b^2|--+\rangle & \rightarrow & a^1b^2|1_H,0_V \rangle _{w_1} |1_H,0_V \rangle _{w_2}|1_H,0_V \rangle _{w_3}  \nonumber\\
a^0b^3|---\rangle & \rightarrow & a^0b^3|1_H,0_V \rangle _{w_1}
|1_H,0_V \rangle _{w_2} |0_H,1_V \rangle _{w_3} \nonumber
\end{eqnarray}
where $a=[\frac{(1+sin\theta)}{2}]$, and
$b=[\frac{(1-sin\theta)}{2}]$. Of course the states appear to be
of different length but as we explained in Section II this is not
the case, the missing modes are occupied by vacuum states (which
carry no information).  The logic of our compression is that the
state with the highest probability (the one that appears most in
the classical language) is encoded in the shortest possible form.
Note that this is different to Schumacher's strategy. Schumacher
only encodes the states in the typical subspace, and all the
other states are deleted (leading to unfaithful compression for
the finite size message). The states in the typical subspace, on
the other hand, are in Schumacher's case all encoded into
codewords of equal length (asymptotically equal to the original
length times the entropy of the signal). In our scheme the
typical subspace does not have exclusive importance, with all the
messages being encoded faithfully. Note finally that the whole
transformation is unitary and therefore can be implemented in
quantum mechanics (we show how to do so in section V).

From this encoding, when say, $\theta = 45^\circ$, we can infer
the entropy of the string as $1.29$ bits/symbol.  This is better
than our expected optimal of $1.8$ bits/symbol and is a
significant improvement on Schumacher codings $1.88$ bits/symbol.
However, as we will see, it is not appropriate and is actually
misleading to directly compare these results without the added
the respective information required from the classical side
channel. The main advantage of this compression method over that
of Schumacher's is that this is lossless in the finite case, i.e.
signal can be completely recovered, unlike in Schumacher's case
where a certain loss in fidelity is inevitable. It is clear that
our example with $3$ qubits can in fact be applied to any number
of qubits (or, more accurately, to quantum systems of any
dimension) by continuing with the principle of encoding less
likely strings into states with more photons. This mapping is
perfectly well defined and unique even given the case that we
have messages of equal probability, where here we can arbitrarily
choose which message to encode into the shorter word.

An important point to make is that in this scheme we no longer
need to use the classical notion of unique decipherability (Fig
3)\cite{cover,jones} for defining codeword mappings. This is
because given the encoding technique any codeword set that
represents a 1-1 map between codeword and letter state is
sufficiently effective in being uniquely decipherable (U.D.).
Therefore the quantum notion of U.D., as directly applied in this
case, is stronger and allows for shorter codewords than is
classically possible, something that has has also been considered
by Bostr\"{o}m \cite{bostroem}.

In terms of decompression, classically we make use of the fact
that we have the length information of each codeword.  However in
quantum mechanics encoding the length information of each
codeword along with the respective codeword is quite impractical,
as a number of authors have pointed out
\cite{bostroem,cleve,schu2}. This is because in quantum mechanics
in order to infer the individual length of a codeword would
require there to be a measurement of some sort and performing any
measurement would collapse this superposition onto any one of
those codewords irreversibly, resulting in a disturbance to the
state and therefore an unacceptable loss of information. It is
therefore indeed fortunate that in order to faithfully decompress
(i.e. replace the redundancy that was removed by quantum
compression) we need only to know the total length of the message
that was initially encoded (i.e. total number of qubits
transmitted) rather than the individual lengths of each codeword
(i.e. length of each letter state). With having only this total
length information, $l_t$ we then know the redundancy we need to
add to the compressed signal (i.e. the signal containing the
statistical properties of the original message) in order to
restore the original message. Clearly if this information is
missing we can only probabilistically achieve faithful
decompression by best guessing the original length of the message
as also pointed out by Bostr\"{o}m \cite{bostroem}. As $l_t$
cannot be measured, it must be known by the sender and sent
additional to the compressed quantum message (see Fig. 4)(via a
classical or quantum side channel) or perhaps agreed upon between
sender and receiver prior to communication.  It is worth
clarifying that classically, $l_t$ is always available to us
regardless of the coding scheme employed, as we can easily make a
measurement on its length without any risk of disturbing the
state.

From Landauer's erasure principle \cite{landauer}, briefly
discussed and applied in section III, it is possible to derive an
lower bound on the efficiency of this compression scheme. We use
the fact that according to Landauer when we erase n units of
information we have to increase the entropy of the environment by
n units. If the entropy increase of the environment is less than
this, that then must imply that there is a suitable amount of
information that was not deleted.

The encoding we use to achieve compression is faithful for any
finite length of message, only if, as mentioned before, together
with the compressed message we send another piece of information.
This could be the total length of the uncompressed message, or
instead, slightly more efficiently, the entropy of the message.
So, if the statistical properties of the message are represented
by $\rho_n$, we could send additional $\log n$ (qu)bits along
with the compressed message to represent the length of the total
signal, or, at best send $\log S(\rho_n)$ (which is $\le \log
n$). Therefore from Landauer's principle we expect that the limit
to compression in our scheme is bounded from below by:

\begin{equation}
S_{1-1}(\rho_n) \ge S(\rho_{n}) - \log(n)\nonumber
\end{equation}
if we are sending log(n) bits of length information or
\begin{eqnarray}
S_{1-1}(\rho_n) \ge S(\rho_{n}) - \log(S(\rho_{n}))\nonumber
\end{eqnarray}
if we, more efficiently, only send the entropy of the total signal, from which it is
possible to infer the length information.

A more rigorous proof to these bounds and to the 1-1 quantum
compression scheme can however be obtained using results from
Cover and Prisco\cite{cheong,prisco}.
From our encoding scheme we can see that the average length of
the {\em i'th} codeword is:
\begin{eqnarray}
l_i= \bigg\lceil \log(\frac{i}{2} + 1) \bigg\rceil \nonumber
\end{eqnarray}
\noindent and therefore by definition \cite{jones}, the average
word length associated with this coding scheme, $L_{1-1}$ is:
\begin{eqnarray}
L_{1-1}= \sum_i^N p_il_i = \sum_i^N p_i \bigg\lceil  \nonumber
\log(\frac{i}{2}+ 1) \bigg\rceil \nonumber
\end{eqnarray}
\noindent In a similar fashion to the Shannon entropy and minimum
average word length for U.D. codes \cite{jones}, we define the
lower bound of our 1-1 average word length as the corresponding
1-1 entropy, $H_{1-1}$.  This 1-1 entropy tells us the best that
we can compress to using 1-1 codes and it is related to the
Shannon entropy in the following manner:
\begin{eqnarray}
H_X - H_{1-1} \leq \sum_{i=1}^N p_i \bigg(\log \frac{1}{p_i} -
\log (\frac{i}{2}+ 1)\bigg) \nonumber
\end{eqnarray}
\noindent and by then using the method of Lagrange multipliers to
maximize the right hand side of the expression as shown by
\cite{cheong} we find that:
\begin{eqnarray}
H_{1-1}(S) \ge H(S) - \log(n) - 3 \nonumber
\end{eqnarray}
\noindent This proof by \cite{cheong} was later refined by
\cite{prisco} to
\begin{eqnarray}
H_{1-1}(S) \ge H(S) - \log(H(S)) - H(S) \log{(1+
\frac{1}{H(S)})}\nonumber
\end{eqnarray}
Given that the 1-1 part of our encoding scheme may be essentially
considered to be classical (since classical mechanics is a
special case of quantum mechanics in the diagonal basis) we can
interchange the Shannon for the von Neumann entropy and obtain an
exact lower bound for the compression of our quantum 1-1 coding:
\begin{eqnarray}
S_{1-1}(\rho_n) \ge S(\rho_n) - \log(S(\rho_n)+1) - S(\rho_n)
\log(1+ \frac{1}{S(\rho_n)})\nonumber
\end{eqnarray}
where $S_{1-1}(\rho_n)$ is our quantum 1-1 entropy and
$S(\rho_n)$ is the entropy of the total (unencoded) message. So
we see that for large $S(\rho_n)$ this bound coincides with the
one obtained independently and more physically through Landauer's
erasure.
Therefore from the 3 qubit example given earlier the total
entropy of the state after compression (i.e. including classical
side channel) is therefore $S(\rho)+log(N) = 1.81$ bits/symbol,
still an improvement on 1.88 bits/symbol by Schumacher. It is the
case however that regardless of the efficiency of this scheme in
the finite limit, both schemes tend towards the same von Neumann
entropy. In summary both these methods could be equally useful in
quantum data compression depending on the required accuracy,
speed and convenience of the compression algorithm. Our motivation
was to optimise total energy which we achieve by having an even
greater permissible codespace (i.e. 1-1 rather than U.D.
codespace) and hence on average shorter codewords available to
us. As mentioned it may be the case that different schemes build
on these ideas to optimise resources other than energy e.g.
compression time, circuit complexity, difficulty of
implementation, equipment availability or cost.

\section{Practical Issues}

In this section we discuss practical issues related to realising
a a very simple instance of our 1-1 quantum encoding scheme. We
will be encoding two quantum bits in the state $|\psi_0\rangle$
or $|\psi_1\rangle$ where as in the earlier 3 qubit example
$|\Psi_0\rangle = \cos (\theta/2) |0\rangle + \sin (\theta/2)
|1\rangle $ and $|\Psi_1\rangle = \sin (\theta/2) |0\rangle +
\cos (\theta/2) |1\rangle$ with $p_0=p_1=1/2$. As in the 3 qubit
example, going into the basis where the density matrix is
diagonal and then mapping the respective letter states to
corresponding codewords in order of most probable to least
probable (again here assuming $a > b$), we get:

\begin{eqnarray}
a^2b^0|++\rangle & \rightarrow & a^2b^0|1_H,0_V \rangle|0_H,0_V \rangle \equiv a^2b^0|1_H,0_V \rangle                         \nonumber\\
a^1b^1|-+\rangle & \rightarrow & a^1b^1|0_H,1_V \rangle|0_H,0_V \rangle \equiv a^1b^1|0_H,1_V \rangle                        \nonumber\\
a^1b^1|+-\rangle & \rightarrow & a^1b^1|1_H,0_V \rangle|0_H,1_V \rangle \equiv a^1b^1|1_H,0_V \rangle|0_H,1_V \rangle         \nonumber\\
a^0b^2|--\rangle & \rightarrow & a^0b^2|0_H,1_V \rangle|0_H,1_V \rangle \equiv
a^0b^2|0_H,1_V \rangle|0_H,1_V \rangle\nonumber
\end{eqnarray}
Note that this operation just tells us to annihilate the second
photon if it is the state $|+\rangle$ and map $|+\rangle$ to
$|1_H, 0_V \rangle$ and $|-\rangle$ to $|0_H, 1_V \rangle$.  So
in order to implement this transformation we clearly need to be
able to perform a conditional operation from the polarisation
degrees of freedom to the spatial degrees of freedom.  The
subsequent transformation is then just a change of basis, from a
$+/-$ basis to a $H/V$ basis, which is known as the Hadamard
transform and is easy to implement. We know that as we have an
orthogonal set on the left hand side and an orthogonal set on the
right hand side, according to quantum mechanics there must be a
unitary transformation to implement this.  Since Hadamard is a
simple transformation to implement we only need to show how to
implement the following two qubit transformation:
\begin{eqnarray}
|1_H,0_V \rangle|1_H,0_V \rangle & \rightarrow & |1_H,0_V \rangle         \nonumber\\
|0_H,1_V \rangle|1_H,0_V \rangle & \rightarrow & |0_H,1_V \rangle          \nonumber\\
|1_H,0_V \rangle|0_H,1_V \rangle & \rightarrow & |1_H,0_V \rangle|0_H,1_V \rangle          \nonumber\\
|0_H,1_V \rangle|0_H,1_V \rangle & \rightarrow & |0_H,1_V \rangle
|0_H,1_V \rangle \nonumber
\end{eqnarray}
which actually amounts to deleting the second photon if its state
is $H$ and otherwise leaving everything the same. Note that, due
to linearity of quantum mechanics, a superposition of states on
the left hand side would be transformed into the corresponding
superposition of the states on the right hand side. This means
that we will have elements of unequal length (different number of
photons) present in the superposition. While this may in practice
be difficult to prepare, there is nothing fundamental to suggest
that in principle such states cannot be prepared, as we show next.

In order to implement this transformation, one possible method is presented in Fig. 5
in the form of a simple quantum computational network. In this circuit, we first need
to distinguish the two modes as we only want to delete a particle from the second (and
not the first) mode. We can imagine that in practice if these are two light modes,
then we actually need to distinguish their frequencies $\omega_1$ and $\omega_2$, and
this could be done by a prism splitting the two frequencies at {\bf A}. Therefore the
two modes now occupy different spatial degrees of freedom. Next we need to distinguish
the two polarisations in the second mode, which in the case of a photon would involve
a polarisation dependent beam splitter (PDBS) at {\bf B}. Now, after this
beam-splitter we can distinguish both the frequency and the polarization in the second
mode, and so we only need to remove a photon from the second mode if we have H
polarization. Here we use the trick we mentioned in the Landauer's erasure section,
namely that we swap the photon in the second mode with an environmental vacuum state
conditional on it being horizontally polarized (and otherwise we do nothing). If
initially we had a superposition of all states $a |1_H,0_V \rangle|1_H,0_V \rangle +
b|0_H,1_V \rangle|1_H,0_V \rangle  + c |1_H,0_V \rangle|0_H,1_V \rangle + d|0_H,1_V
\rangle|0_H,1_V \rangle$, and the state of the environment was $|0_H,0_V\rangle$, then
after the swap, the state will be
\begin{eqnarray}
& a &  |1_H,0_V \rangle|0_H,0_V\rangle|1_H,0_V \rangle + b|0_H,1_V
\rangle|0_H,0_V\rangle|1_H,0_V \rangle + \nonumber \\
& c & |1_H,0_V \rangle|0_H,1_V \rangle|0_H,0_V\rangle + d|0_H,1_V \rangle|0_H,1_V
\rangle|0_H,0_V\rangle \nonumber
\end{eqnarray}
We now need to perform a simple Hadamard transformation on the environment such that
\begin{eqnarray}
|1_H, 0_V\rangle &\rightarrow & |0_H, 0_V\rangle + |1_H, 0_V\rangle \nonumber \\
|0_H, 0_V\rangle &\rightarrow & |0_H, 0_V\rangle - |1_H, 0_V\rangle \nonumber
\end{eqnarray}
after which the total state can be written as
\begin{eqnarray}
(& a & |1_H,0_V \rangle|1_H,0_V \rangle + \ldots d|0_H,1_V \rangle|0_H,1_V \rangle) |0_H,0_V\rangle + \nonumber\\
(& a & |1_H,0_V \rangle|1_H,0_V \rangle + \ldots - d|0_H,1_V \rangle|0_H,1_V \rangle)
|1_H,0_V\rangle \nonumber
\end{eqnarray}
Note that after performing a measurement on the environment (at {\bf C}), if we obtain
the outcome $|0_H,0_V\rangle$, then the resulting state of two photons is already our
encoded state, while if the outcome is $|1_H,0_V\rangle$, then the state is the
encoded state up to a negative phase shift in the last two elements of the
superposition. This can be corrected by applying a simple phase shift conditional on
the second photon being vertical. The whole operation at {\bf C} can also be performed
coherently without performing the measurement as indicated in Fig 5. We acknowledge
that this operation may not be simple to execute in practice and may require a $100$
percent effective photo-detection scheme which is currently unavailable. However, this
gate can certainly be implemented with some probability at present. At the end, we
need to reverse the operation of the PDBS, and then reverse the operation of the
prism, thus finally recombining the two modes into the same spatial degree of freedom.
The resulting state is our encoded state and can then be sent as such.

\section{Discussion}
In this paper a new variable length quantum data compression
scheme has been outlined. By looking at quantum data compression
in the second quantisation framework, we can generate variable
length codes in a natural and efficient manner without having the
significant memory overhead common to other variable length
schemes \cite{bostroem}. The quantum part of our signal is
compressed beyond the von Neumann limit, but at the expense of
having to communicate a certain amount of classical information.
By sending the total length of the transmitted signal through a
classical channel enables us to compress and decompress with
perfect fidelity for any number of qubits.  We have presented an
argument based on Landauer's erasure principle which provides us
with a with a lower bound on the efficiency of our compression
scheme. This is independently verified by classical results due
to Cover \cite{cheong} and Prisco \cite{prisco}. As expected, the
sum of the classical and the quantum parts of the compressed
message still tends towards the limit given by the von Neumman
entropy. Asymptotically, the quantum part dominates over the
classical part and becomes equal to the von Neumman entropy. The
tightest compression bound for our scheme is not known.

Note that we assume that both the sender and receiver know
exactly the properties of the source, i.e. they know the quantum
states the source emits and the corresponding probabilities and
modes within which they are emitted. This of course means that our
scheme is not universal.  It is unlikely however that in a
universal scheme the sender and receiver would need less
information than this to perform compression, e.g. just knowledge
of the entropy of the source and length of message without
knowledge of the density matrix (or perhaps even this is
unnecessary \cite{caroline}). But this is a separate issue that
warrants further investigation.

Our encoding has a novel feature that it involves superpositions
of different numbers of photons within the superposition states.
We acknowledge that there may be a ``superselection" rule that
prohibits the nature of this approach. However, we believe that,
while these states might be difficult to prepare, they are
certainly not impossible according to the basic rules of quantum
mechanics. To support this we offer a general way of implementing
our scheme in the simple case of encoding two quantum bits.
Whether the space-time complexity of our implementation is most
efficient in practice remains an open question.

As it is, our encoding is a unitary transformation and the
receiver applies the decoding operation (inverse of the unitary
transformation) to decompress the quantum message.  In the case
that we encode and send classical bits, the receiver may wish to
infer the original classical string that was sent. The receiver
can then perform measurements on the decompressed quantum states
to infer the original classical letters.  Since the original
classical letters are by definition fully distinguishable and if
the transmitted quantum states are orthogonal only then can this
final step be done with perfect efficiency (this is of course a
special case of our most general quantum scheme).

It is also worth noting that it is on the sequence of these
quantum states i.e. on the total message, that our compression
scheme acts. This means that this scheme would not be so useful
in an application where instantaneous lossless decompression is
required, where one would have to wait for all the photons to
arrive before beginning the decoding operation.  In the event
that the receiver starts the decompression operation in advance
of the last photon arriving, he truncates the signal and hence
will not be able to decode the original signal with perfect
fidelity.

In our scheme it is the average length of the message, or more
appropriately the average energy required to represent the
information within the quantum state, that tends in the
asymptotic limit towards the von Neumann entropy. We therefore
decided instead to re-formulate compression from an energy
perspective, as the measure is then more consistent as an optimal
measure of a systems information carrying ability. As we are
aware in quantum mechanics, energy and information are
intricately linked, far more so than photon number and
information.  We are interested in primarily reducing the energy
required to represent the message, which we stress is not
affected by the fact that we need to wait until the whole signal
is received.  In our framework the incorporation of any vacuum
states to extend the variable length message to the same length
as the longest component of the superposition, by definition,
does not increase the energy total for the message.  In reality
of course we do not even have to wait for the whole signal, we can
just truncate it at the average length of the signal and although
we end up with a lossy compression scheme we still tends towards
the von Neumann entropy asymptotically.  The issue of waiting
until the whole signal arrives really is to do with the fact that
we cannot measure the length of the signal without collapsing it
into a particular length, which is not what we want as we want to
keep intact the superposition and consequently preserve the rest
of the information within the system.

Our approach raises a number of interesting questions. Firstly,
it gives us a more physical model of data compression and relates
the entropy to the minimum energy required to represent the
information contained within a quantum state. This could be very
useful from an energy saving perspective and gives a guideline as
to the minimum temperature we could cool a system to before we
begin to loose information. Another benefit to this compression
scheme is that it does not depend on the nature of particles, the
scheme applies equally well to both bosonic and fermionic
systems. The reason for this is that we never put more than one
particle per state when we are encoding and therefore we never
need to consider the Pauli exclusion principle. Whether this
principle plays a more important role in data compression, i.e.
whether there could be a fundamental difference between the
bosonic and fermionic systems ability to store (and in general
process) information is not yet known. The ultimate bound due to
Bekenstein \cite{bekenstein} suggests that the answer is ``no",
however, specific encodings may highlight differences between the
two kinds of particles.

Finally, our scheme assumes that the encoding and the decoding processes as well as
the possible channel in between the two are error free. In practice this is, of
course, never true and it is interesting to analyze to what extent our scheme suffers
in the presence of noise and decoherence at its various stages. We hope that our work
will stimulate more research into quantum data compression as well as experimental
realization in the optical and the solid state domain.

\section{Acknowledgements}
We would like to acknowledge useful discussions with K.
Bostr\"{o}m, D. Bouwmeester, G. Bowen, C. Rogers and useful
communication with T. Cover and J. Keiffer.  L. R. acknowledges
financial support from Invensys plc. V.V. would like to thank
Hewlett-Packard, Elsag s.p.a. and the European Union project
EQUIP (contract IST-1999-11053) for financial support.

\begin{figure}[ht]
\begin{center}
\hspace{0mm} \epsfxsize=5.6cm
\epsfbox{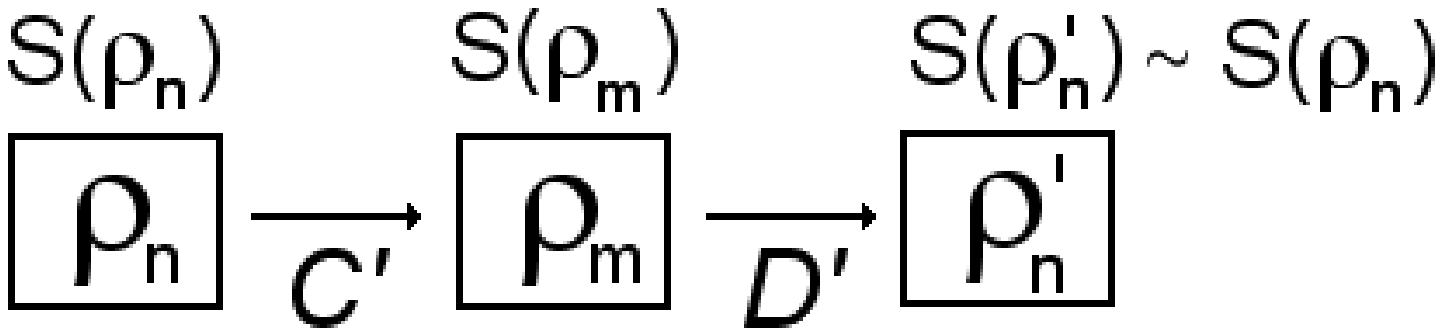}\\
\begin{caption}{\label{fig:schumcoding} In Schumacher's coding the first operation
${C}'$ compresses a quantum source $\protect\rho_n$ stored in $n$
qubits into $m = nS\left( \protect\rho_n \right)$ qubits. This is
then decompressed by an operation $D'$, and for a finite length
message the output state is not in general the same as input. In
the asymptotic limit, on the other hand, the source is accurately
recovered.}
\end{caption}
\end{center}
\end{figure}
\begin{figure}[ht]
\begin{center}
\hspace{0mm} \epsfxsize=7.5cm
\epsfbox{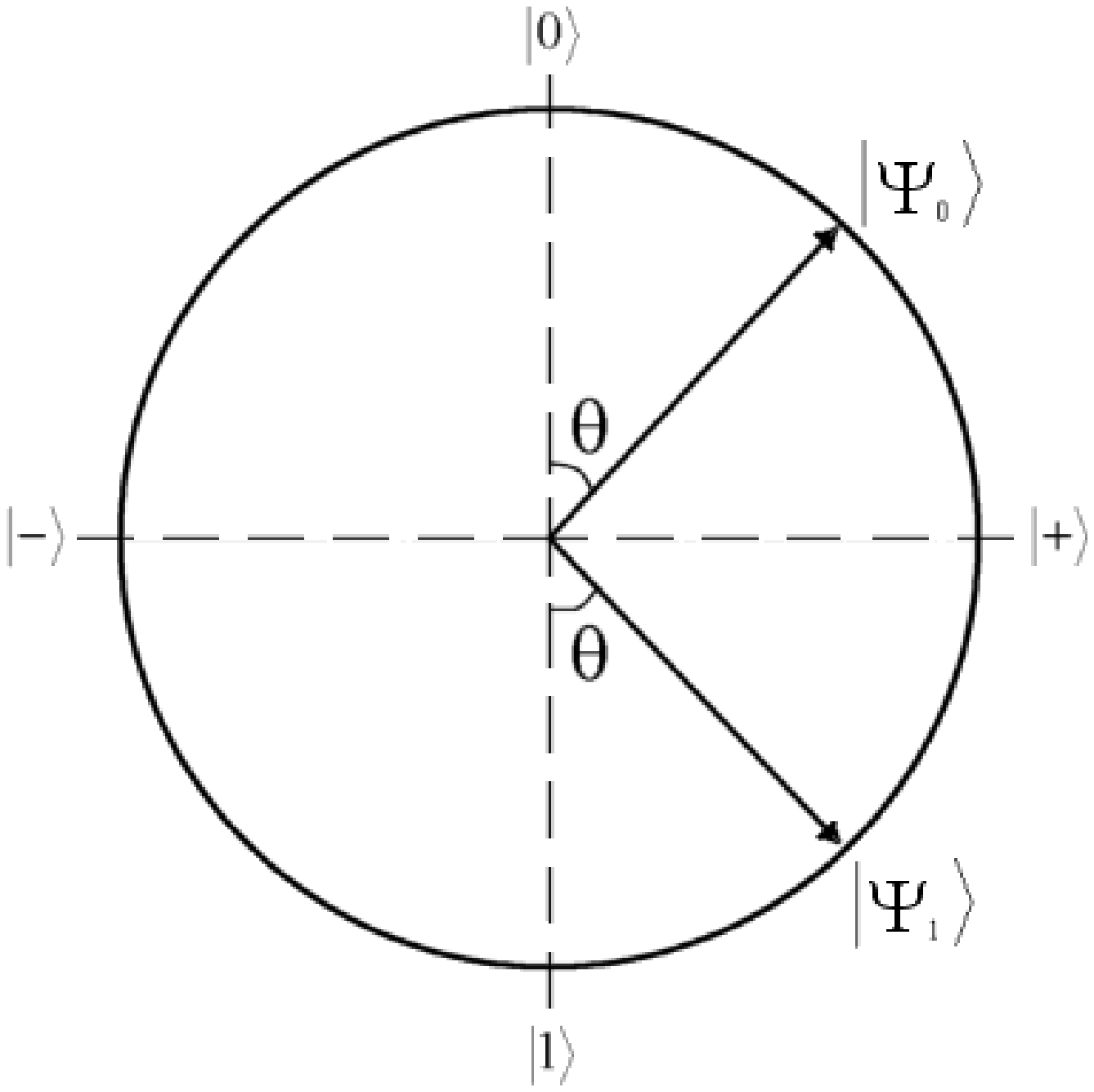}\\
\begin{caption}{\label{fig:bloch} This figure shows the two
non-orthogonal states on the Bloch sphere which are used to
encode the message.}
\end{caption}
\end{center}
\end{figure}
\begin{figure}[ht]
\begin{center}
\hspace{0mm} \epsfxsize=7.5cm
\epsfbox{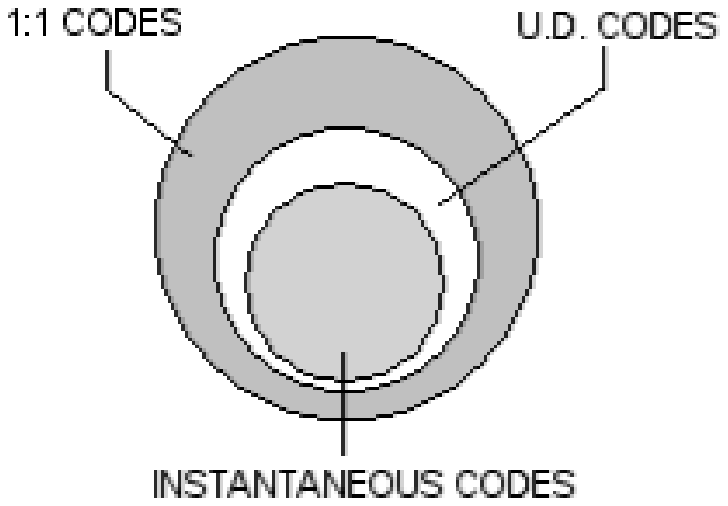}\\
\begin{caption}{\label{fig:codespace} Illustration of Codespace: Uniquely
decipherable (U.D.) and prefix-free (instantaneous) codes are a
subset of the 1-1 codes used in our data compression scheme.
Classically, 1-1 codes are not very useful for data compression
as they usually require another symbol signaling the end of one
letter and the beginning of another one. However, our presented
quantum scheme enables us to make us of 1-1 codes in a way that
is not classically practical.}
\end{caption}
\end{center}
\end{figure}
\begin{figure}[ht]
\begin{center}
\hspace{0mm} \epsfxsize=7.5cm
\epsfbox{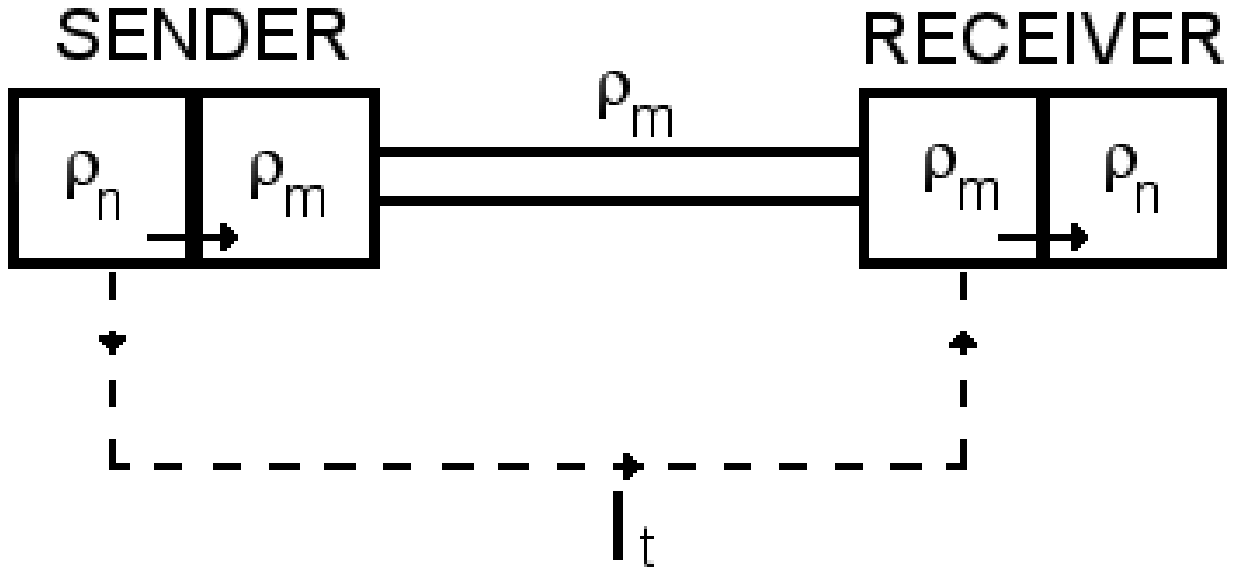}\\
\begin{caption}{\label{fig:algoverview} This diagram presents the core of our
quantum 1-1 compression algorithm. Initially, $\rho_n$ is
compressed into $\rho_m$ and sent together with the classical
message, $l_t$, containing the information about the total number
of input qubits (i.e. total length of the signal). On
decompression, using the information in $l_t$, the original signal
is faithfully recovered for any number of qubits.}
\end{caption}
\end{center}
\end{figure}
\begin{figure}[ht]
\begin{center}
\hspace{0mm} \epsfxsize=7.5cm
\epsfbox{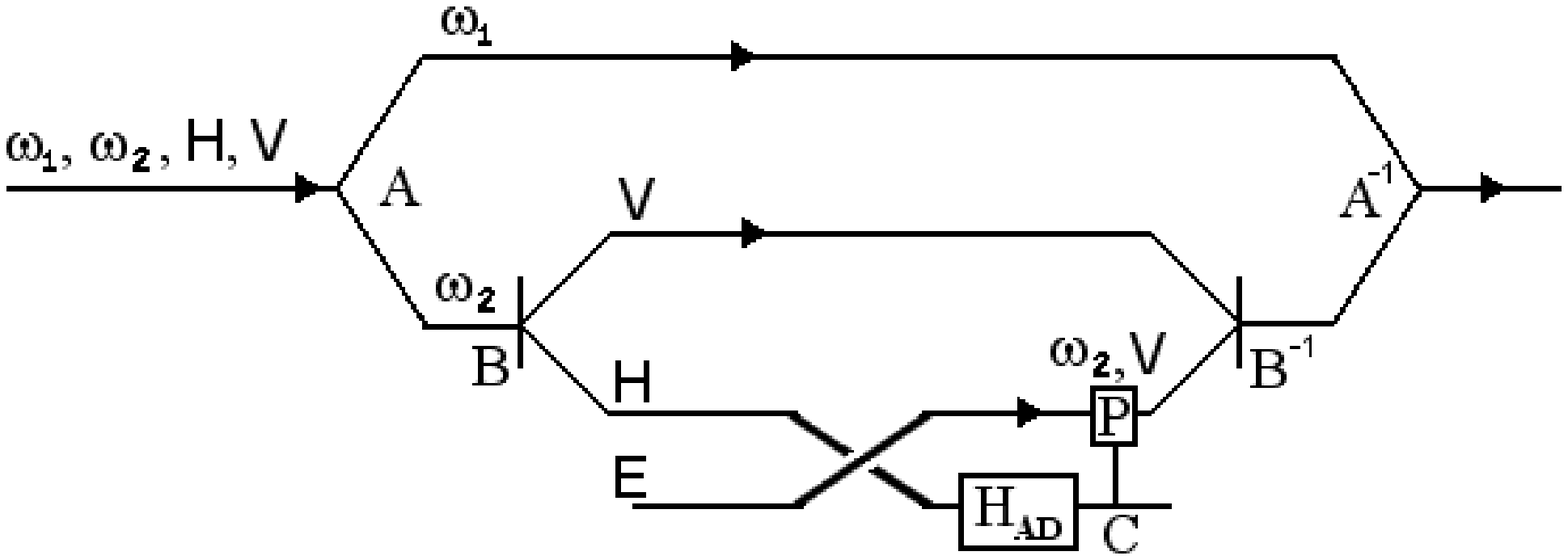}\\
\begin{caption}{\label{fig:experiment} A circuit for our quantum data compression scheme applied
to $2$ qubits. The two photons $\omega_1$ and $\omega_2$ are
initially split according to their frequency, after which the
photon in frequency $\omega_2$ is further split according to its
polarization. The H branch is then swapped with an environmental
vacuum state, while nothing happens to the V branch of
$\omega_2$. The gate $H_{AD}$ is the Hadamard transform
subsequently acting on the environment as defined in the text.
The following gate is a conditional phase gate, $P$, introducing
a negative phase in the second photon $\omega_2$, only if its
polarization is V, and the environment has one photon. At the end
the two polarizations and then the two photons are recombined
back into the single spatial degree of freedom. Our circuit is
completely general and could be applied to different kinds of
particles, such as electrons for example.}
\end{caption}
\end{center}
\end{figure}
\end{document}